\documentclass[superscriptaddress,twocolumn]{revtex4}
\usepackage{epsfig,amsmath,graphicx,amssymb}
\usepackage[colorlinks=fals,linkcolor=blue]{hyperref}
\def\be{\begin{equation}}
\def\ee{\end{equation}}
\def\bea{\begin{eqnarray}}
\def\eea{\end{eqnarray}}
\begin{document}
\title{Discontinuity of curvature on axisymmetric Willmore surfaces}
\author{Xiaohua Zhou}
\email{zhouxiaohua@xijing.edu.cn; xhzhou08@gmail.com}
 \affiliation{Department of Physics, Xijing University, Xi'an 710123, China}
 \affiliation{Engineering Technology Research Center of Controllable Neutron Source and Its Application, Xijing University, Xi'an, 710123, China}

\date{\today}
\begin{abstract}

The equilibrium shapes of vesicles are governed by the general shape equation which is derived from the minimization of the Helfrich elastic free energy and can be reduced to the Willmore equation in a special case. The general shape equation is a high order nonlinear partial differential equation and it is very difficult to find analytical solution even in axisymmetric case, which is reduced to a seconder ordinary differential equation. Traditional axisymmetric shape equation is with the turning radius as the variable. Here we study the shape equation with the tangential angle as the variable. In this case, the Willmore equation is reduced to the Bernoulli differential equation and the general solution is obtained conveniently. We find that the curvature in this solution is discontinuous in some cases, which was ignored by previous researchers. This solution can satisfy the boundary conditions for open vesicle with free edges.
\emph{}

\pacs{XX, XX}

\textbf{keywords:} {Vesicle, solution, boundary conditions}
\end{abstract}

 \maketitle

 \section{introduction}
Biological membranes are composed of two layers of phospholipid molecules.
In aqueous environment, phospholipid bilayer can forme closed vesicles \cite{Szoka1980, Evans1990PRL, Hotani1984JMB} as well as open vesicles \cite{Saitoh1998PNAS}. By Comparing lipid bilayer with liquid crystals, Helfrich pointed out that membranes can be taken as two-dimensional (2D) surfaces and their equilibrium shapes are determined by minimization of the elastica free energy \cite{Helfrich1973}
\bea\label{Energy}
E=\int\int[\frac{k_{c}}{2}(2H+C_{0})^{2}+k_g\Lambda] dA,
\eea
where $H$, $\Lambda$ and $C_0$ are the mean curvature, Gaussian
curvature and spontaneous curvature of the surface, respectively. $dA$ is the
surface area element, $k_{c}$ and $k_g$ are the bending moduli of the membrane. For
close vesicles, considering the surface area and volume constraints, the energy functional
is
\bea\label{Energy Functional}
F=E+\lambda \int\int dA+ p\oint dV,
\eea
where $\lambda$ is the surface tension coefficient, $p$ in the pressure difference between the inner and out of the vesicle and $dV$ is the volume element of the vesicle volume. By studying the
first variation of $\delta^{(1)}F=0$, Ou-yang and
Helfrich obtained the general shape equation of closed
membranes \cite{OuYang1987PRL, OuYang1989PRA}
\bea\label{general equation}
\nonumber
              k_c(2H+C_0)(2H^2-2\Lambda-C_0H)
    -2\lambda H \\+2k_c\nabla^2H+p = 0,
\eea
where $\nabla^{2}$ is the Laplace-Beltrami operator in 2D case.
The above equation is a high order nonlinear partial differential equation and it is very difficult to find analytical solution even in axisymmetric, which can be reduced to a seconder ordinary differential equation.
In past two decades, several special analytical solutions have been found, such as the Clifford torus \cite{OuYang1990PRA}, discounts \cite{Naito1993PRE}, the
beyond-Delaunay surface \cite{Naito1995PRL} and the general solution for cylindrical case. All of these solutions are shown in Ref.\cite{OuYangBook}. Besides, numeric solutions in axisymmetric case have been studied extensively and the phase diagram is obtained \cite{Seifert1991PRE}. Many interesting non-axisymmetric solutions are obtained by using the finite elemental method \cite{YanJ1998PRE,ZhouXH2008IJMPB}. 
 
 Moreover, the equilibrium shape equations for multi-component vesicles, open vesicles with free boundary as will as vesicle adhesion structures are investigated and the corresponding shapes are obtained \cite{ DuQ2008JMB, TuZC2010JCP, Seifert1990PRA, Deserno2007PRE, DuQ2010PRE}. In these models, Eq.~(\ref{general equation}) also needs to be satisfied (It needs $p=0$ for open vesicles). Therefor, to find new analytic solutions for Eq.~(\ref{general equation}) is an attractive and challenging work \cite{TuZC2013CPB}.

Specially, if $\lambda=p=C_0=0$, Eq.~(\ref{general equation}) is reduced to the well-known Willmore equation which is used to describe the equilibrium configurations of thin elastic shell \cite{Poisson1833}. Recently, some researchers found an analytical solution for the Willmore equation in the axisymmetric case \cite{Vassilev2014AIP, TuZC2016arXiv}. It inspires us to find new solutions for the equilibrium shape equation (\ref{general equation}).
Traditionally, the axisymmetric shape equation for vesicle is with the turning radius as the variable \cite{HuJG1993PRE,ZhengWM1993PRE}.
Here we give the shape equation with the tangential angle as the variable. In this case, the Willmore equation is reduced to the Bernoulli differential equation and the general solution is obtained, which is nothing but the new solution recent works \cite{Vassilev2014AIP, TuZC2016arXiv}. Meanwhile, we find that the curvature in this solution is discontinuous in some cases, which is ignored by previous researchers.
We also point out that this solution can satisfy the boundary conditions for open vesicle with free edges.

\section{Axisymmetrical shape equation and a special solution}

For a axisymmetric surface, let the generating line be around the $z$-axis and $\rho$ be the turning radius, there are
\bea\label{xyz}
x=\rho \cos\phi, y=\rho \sin\phi, z=\int\tan\psi(\rho)d\rho,
\eea
where $\phi$ is the azimuthal angle and $\psi$ is the tangent angle of the profile curve.
The surface can be expressed as
\bea\label{r}
 \textbf{r}=\{\rho \cos\phi,\rho \sin\phi, z \} .
\eea
Defining $()'=\frac{d()}{d\rho}$, we have
\bea\label{H1}
 &&2H=\frac{\sin\psi}{\rho}+(\sin\psi)^\prime,~~\Lambda=\frac{\sin\psi}{\rho}(\sin\psi)^\prime,\\ &&\nabla^2=\frac{\cos\psi}{\rho}\bigg[\frac{\partial}{\partial\rho}\big(\rho\cos\psi\frac{\partial}{\partial\rho}\big)
 +\frac{\partial}{\partial\phi}\big(\frac{\sec\psi}{\rho}\frac{\partial}{\partial\phi}\big)\bigg]. %
\eea
Then the general shape
equation is reduced to \cite{ZhengWM1993PRE}
\bea\label{Eq1}
\nonumber 2 H^\prime\cos\psi+(2H-C_0)\psi^\prime\sin \psi-\tilde{\lambda} \tan\psi\\
+\frac{\eta-\tilde{p}\rho^2}{2\rho \cos\psi}-\frac{\tan\psi}{2}(2H-C_0)^2=0
\eea
where $\eta$ is an integral constant, $\tilde{\lambda}=\lambda/k_c$ and $\tilde{p}=p/k_c$. The above is a seconder order nonlinear differential equation with the variable $\rho$. Although it is very difficult to be solved generally, several analytic solutions were found \cite{OuYangBook}.

Now, let us consider another way to express this equation by using the tangential angle $\psi$ as the variable.
Defining $\dot{\rho}=\frac{d\rho}{d\psi}$, we have $()'=\frac{d()}{d\rho}=\frac{d()}{d\psi}\frac{d\psi}{d\rho}=\frac{1}{\dot{\rho}}\frac{d()}{d\psi}$ and
\bea\label{H2}
 2H=\frac{\sin\psi}{\rho}+\frac{\cos\psi}{\dot{\rho}},~~\Lambda=\frac{\sin\psi\cos\psi}{\dot{\rho}\rho},\\ \nabla^2=\frac{\cos\psi}{\rho}\bigg[\frac{1}{\dot{\rho}}\frac{\partial}{\partial\psi}\big(\frac{\rho}{\dot{\rho}}\cos\psi\frac{\partial}{\partial\psi}\big)
 +\frac{\partial}{\partial\phi}\big(\frac{\sec\psi}{\rho}\frac{\partial}{\partial\phi}\big)\bigg].
\eea
The shape equation is changed to
\bea\label{Eq2}
\nonumber \frac{2\dot{H}}{\dot{\rho}}\cos\psi+\frac{1}{\dot{\rho}}(2H-C_0)\sin \psi-\tilde{\lambda} \tan\psi\\+\frac{\eta-\tilde{p}\rho^2}{2\rho \cos\psi}-\frac{\tan\psi}{2}(2H-C_0)^2=0.
\eea
A trial solution can be defined as $\rho=\exp[f(\psi)]$. Substituting it into above equation we attain
\bea\label{Eq3}
\nonumber\big[2 \tilde{p} e^{3f}\sec\psi-2e^{2f}(C_0^2+2\tilde{\lambda})\tan\psi+2e^{f}(\eta\sec\psi\\
\nonumber+2C_0\sin\psi\tan\psi)-(3+\cos2\psi)\tan\psi\big]\dot{f}^3\\
-4\cos^2\psi \ddot{f}-\sin2\psi \dot{f}=0.
\eea
If $\tilde{p}=\tilde{\lambda}=C_0=\eta=0$, the above equation should be the axisymmetric Willmore equation. Meantime, it is reduced to the following Bernoulli differential equation
\bea\label{Eq4}
(3+\cos2\psi)\tan\psi\dot{f}^3+4\cos^2\psi \ddot{f}+\sin2\psi \dot{f}=0.
\eea
This equation has following general solution
\bea\label{S1}
\nonumber \dot{f}^{-2}&=&-2e^{2\int{P(\psi) d\psi}}\bigg[\int Q(\psi)e^{-2\int{P(\psi) d\psi}} d\psi-\frac{I}{2}\bigg]\\
            &=&\tan^2\psi+I \sec\psi,
\eea
where $P(\psi)=\frac{1}{2}\tan\psi$, $Q(\psi)=-\frac{(3+\cos2\psi)\sin\psi}{4\cos^3\psi}$ and $I$ is an integral constant. Then we obtain
\bea\label{S2}
\rho=\rho_0\exp\bigg(\pm\int\frac{ d \psi}{\sqrt{\tan^2\psi+I \sec\psi}}\bigg),
\eea
where $\rho_0$ is an integral constant. This solution was obtained by using the Noether theorem in a recent work \cite{TuZC2016arXiv}. Considering $\tan\psi=\frac{dz}{d\rho}$, we obtain
\bea\label{dz}
\frac{dz}{d\psi}=\frac{dz}{d\rho} \frac{d\rho}{d\psi}=\frac{\pm \rho}{\sqrt{\tan^2\psi+I \sec\psi}}\tan\psi.
\eea
 Making use of $\frac{dz}{d\rho}=\frac{1}{\dot{\rho}}\frac{dz}{d\psi}$ and $\frac{d^2z}{d\rho^2}=\frac{1}{\dot{\rho}^2}\frac{d^2z}{d\psi^2}-\frac{\ddot{\rho}}{\dot{\rho}^3}\frac{dz}{d\psi}$, the above equation is changed to the following Willmore equation \cite{Vassilev2014AIP}
\bea\label{ddz}
\frac{d^2z}{d\rho^2}=\pm\frac{1}{\rho}\bigg[\big(\frac{dz}{d\rho}\big)^2+1\bigg]\sqrt{\big(\frac{dz}{d\rho}\big)^2
+I\sqrt{\big(\frac{dz}{d\rho}\big)^2+1}}.
\eea
By solving this equation, one can get Willmore surfaces. Several example shapes were shown in Refs.~\cite{Vassilev2014AIP, TuZC2016arXiv}. However, we will see that the curvature on these shapes is discontinuous in certain cases.

\section{Discontinuity of curvature on axisymmetric Willmore surfaces }

\begin{figure}[h]\centering
\includegraphics[height=6.5cm]{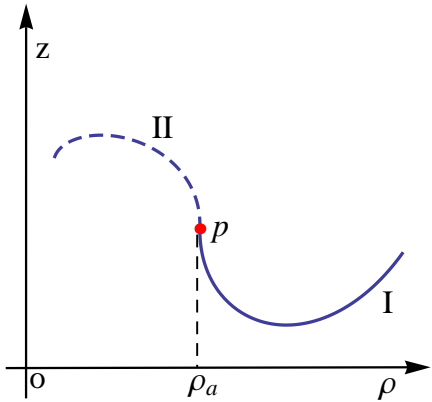}
\caption{A concave curve (part I) and a convex curve (part II) connect at point $p$. }
\label{fig1}
\end{figure}

In Eq.~(\ref{ddz}), the ``$\pm$" indicates that the solution has two parts. Defining
\bea\label{Y}
Y=\frac{1}{\rho}\bigg[\big(z'\big)^2+1\bigg]\sqrt{\big(z'\big)^2
+I\sqrt{\big(z'\big)^2+1}},
\eea
for part I we have $z=z_1(\rho)$ and $z_1''=Y$ and for part II we have $z=z_2(\rho)$ and $z_2''=-Y$. Because $Y\geq 0$, we have $z_1''\geq 0$ and $z_2''\leq 0$. This result determines the concave and convex configurations of $z_1(\rho)$ and $z_2(\rho)$. An example is shown in Fig.~\ref{fig1}, where part I is concave and part II is convex. Refs.~\cite{Vassilev2014AIP, TuZC2016arXiv} shows several shapes with concave and convex parts in a solution. According to our analysis, each shape is possibly the combination of the $z_1(\rho)$ and $z_2(\rho)$ parts. If so, the continuity of this solution on the combination point needs to be discussed.
Making use of Eqs.~(\ref{H2}) and (\ref{Y}), the mean curvature $H$ and $H'$ are
\bea\label{H1H2}
H_1=\frac{z'\big[\big(z'\big)^2+1\big]+Y\rho}{2\rho\big[\big(z'\big)^2+1\big]^{3/2}}, H_2=\frac{z'\big[\big(z'\big)^2+1\big]-Y\rho}{2\rho\big[\big(z'\big)^2+1\big]^{3/2}}, \\
H_1'=H_2'=-\frac{I z'}{4\rho^2}.~~
\eea
where the foot note 1 and 2 are corresponding to part I and part II, respectively. Supposing that the combination point satisfies $\rho=\rho_a$, shape equation (\ref{Eq1}) needs $z(\rho)$, $H(\rho)$, $H'(\rho)$ and $\psi'(\rho)=z''/(1+(z')^2)$ are continuous at $\rho=\rho_a$.

\begin{figure} \centering
\includegraphics[height=6.5cm]{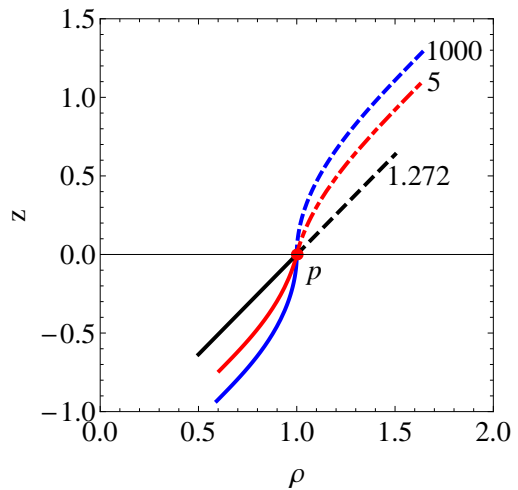}
\caption{Profile curves by solving Eq.~(\ref{ddz}) with $I=-1$, $z(1)=0$ and different $z'(1)$. The curves have a curvature singularity at the point $\rho=1$ except the strait line with $z'(1)=1.272$. }
\label{fig2}
\end{figure}

 If $z_1'(\rho_a)=z_2'(\rho_a)$, we can see $H'$ is continuous, while we cannot make sure $H$ is continuous. Let $H_1=H_2$, we have
\bea\label{dzs}
\big(z'\big)^2+I\sqrt{\big(z'\big)^2+1}=0.
\eea
If $I>0$, there is no real solution for $z'$. Therefor we can not find a point to combine the concave and convex parts in a solution in this case. So, the corresponding shapes in Refs.~\cite{Vassilev2014AIP, TuZC2016arXiv} have discontinuous mean curvature (the Gaussian curvature also is discontinuous). But if $I<0$, that is possible. Then let $z'(\rho_1)$ and $z'(\rho_2)$ be two real solutions of Eq.~(\ref{dzs}), we have
\bea\label{dz1dz2}
z'(\rho_1)=-z'(\rho_2)=\frac{1}{\sqrt{2}}\sqrt{I^2-I\sqrt{4+I^2}}.
\eea
It is not difficult to find that $H$, $H'$ and $\psi'$ are continuous at $\rho=\rho_i$ ($i=1,2$). Therefor, if the connect point between the concave and convex parts is at $\rho=\rho_i$, the combined shape is a solution for Eq.~(\ref{Eq1}) without curvature singularity. However, by solving Eq.~(\ref{ddz}) with condition (\ref{dz1dz2}), we find there is nothing but the straight line solution, which gives the conical surface in 3D case. But it is not an exact solution for Eq.~(\ref{Eq1}). Fig.~\ref{fig2} shows several example shapes with $I=-1$, $z(1)=0$ and different $z'(1)$. By solving Eq.~(\ref{dzs}) we get $z'=\pm 1.272$. Choosing $z'(1)=1.272$, we obtain the straight line in Fig.~\ref{fig2}. Increasing $z'(1)$ will give curved lines with a curvature singularity. For each curve, part I is concave and part II is convex. Similar results can be obtained by choosing negative $z'(1)$.
From Fig.~\ref{fig2} we can see that the solution has asymptotic behaviors. The $z'=z'(\rho_1)$ (or $z'=z'(\rho_2)$) and the $z$-axis are two asymptotic lines. Therefor, the solution of Willmore equation (\ref{ddz}) with $I\neq0$ is either concave or convex shape. The shapes in previous works contain concave and convex configurations in one shape cannot be a solution for Eq.~(\ref{ddz}), because there is a curvature singularity.

\section{Open vesicle with free edges}
\begin{figure}\centering
\includegraphics[height=6.5cm]{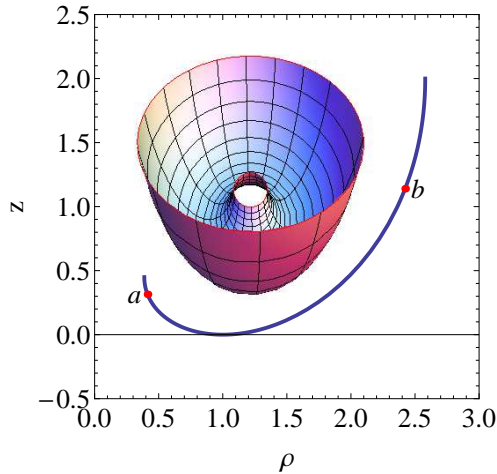}
\caption{ Open shape with two free edges. The real line is the profile curve. Points $a$ and $b$ satisfy the boundary conditions and the inside shape is in 3D case with this two boundaries.}
\label{fig3}
\end{figure}

Experiments have revealed that there are open vesicles with free edges and the corresponding shape equations have been discussed \cite{TuZC2003PRE,TuZC2010JCP}. For open vesicles, the total free energy is
\bea\label{Fopen}
F=\frac{k_c}{2}\int(2H+C_0)^2 dA+\frac{k_g}{2}\int\Lambda dA+\lambda A+\gamma L,
\eea
where $\gamma$ is the line tension of the boundary. In axisymmetric case, $\delta F=0$ yields
the shape equation (\ref{Eq1}) with $p=0$. Besides, the edges should satisfy the following
boundary conditions
\bea
\label{condition1-1}2H-C_0+ \tilde{k} \frac{\sin\psi}{\rho}&=&0,\\
\label{condition1-2}-\sigma\cos\psi \frac{2\dot{H}}{\dot{\rho}}+ \tilde{\gamma}\frac{\sin\psi}{\rho}&=&0,\\
\label{condition1-3}\frac{1}{2}(2H-C_0)^2+\tilde{k}\Lambda+\tilde{\lambda}-\sigma\tilde{\gamma}\frac{\cos\psi}{\rho}&=&0,
\eea
where $\tilde{\gamma}=\gamma/k_c$ and $\tilde{k}=k_g/k_c$. $\sigma=1$ or $\sigma=-1$ if the tangent of the boundary
curve is parallel or antiparallel to the rotation direction, respectively. Between these conditions and Eq.~(\ref{Eq1}), there is the following compatibility condition \cite{TuZC2010JCP}
\bea\label{eta}
\eta=0.
\eea
This condition definitely eliminates the possibility to find equilibrium lipid membranes which is part of known axisymmetric solution of closed vesicles \cite{TuZC2010JCP}. However, the solution in Eq.~(\ref{S2}) gives $\eta=0$, which implies it is a possible solution for open vesicles. When $C_0=\tilde{\lambda}=0$, conditions (\ref{condition1-1}) and (\ref{condition1-2}) give
\bea
\label{condition2-1}\tilde{k}&=&-1\mp\cot\psi\sqrt{\tan^2\psi+I \sec\psi},\\
\label{condition2-2}\tilde{\gamma}\rho &=&-\frac{1}{2}\sigma I .
\eea
Considering that $\gamma$ is the line tension coefficient on the one boundary edge, the total line tension is $T=2\pi\rho\gamma=-\sigma\pi k_c I$, which is in accordance with the result in Ref.~\cite{TuZC2016arXiv}. We note that this solution has two nature free edges, so it cannot describe the shape of closed vesicles. But we can choose two edges on it to satisfy the conditions (\ref{condition2-1}) and (\ref{condition2-2}). On one edge we let $\psi=\psi_1$, $\rho=\rho_1$, $\tilde{\gamma}=\tilde{\gamma}_1$ and $\sigma=\sigma_1$. On another edge we define $\psi=\psi_1$, $\rho=\rho_1$, $\tilde{\gamma}=\tilde{\gamma}_1$ and $\sigma=\sigma_2$. Then conditions (\ref{condition2-1}) and (\ref{condition2-2}) yield
\bea
\nonumber\cot\psi_1\sqrt{\tan^2\psi_1+I \sec\psi_1}
\label{condition3-1}-\cot\psi_2\sqrt{\tan^2\psi_2+I \sec\psi_2}\\=0,~~\\
\label{condition3-2}\sigma_1\tilde{\gamma}_1\rho_1-\sigma_2\tilde{\gamma}_2\rho_2=0.~~
\eea
Eq.~(\ref{condition3-1}) gives $\psi_1=\pm\psi_2$. Solving the above equations we obtain open vesicle with free edges. Fig.~\ref{fig3} gives an example by choosing $\rho_0=1, I=1$ and $\tilde{k}=0.19$ in Eq.~\ref{S2}. For point $b$ there are $\psi_1=1.2, \rho_1=0.41, \sigma_1=1, \tilde{\gamma}_1=-1.21$ and for point $a$ there are $ \psi_2=-1.2, \rho_2=2.42, \sigma_2=1, \tilde{\gamma}_2=-0.21$.

According to the former analysis, when $I=1>0$ the shape is concave or convex. Fig.~\ref{fig3} satisfies this condition.
Moreover, when $I<0$, we find the tangent angle $\psi=\psi(\rho)$ monotonously depends on the change of $\rho$, such as the shapes in Fig.~\ref{fig2}.
In this case, we only can find one boundary angel $\psi$ for condition (\ref{condition3-1}). Considering each solution for Eq.~(\ref{ddz}) with $I\neq0$ has two boundaries, there should be a free boundary which can not satisfy the condition (\ref{condition3-1}).

\section{Conclusions} \label{Conclusions}

In summary, the axisymmetric Willmore equation and its solution is investigated in this study. By choosing a suitable trial solution, this equation can be changed to the Bernoulli differential equation and the general solution can be obtained conveniently. Although this solution has been obtain by previous researchers, the discontinuity of curvature are ignored. Through this study, we know that the shape of this solution either concave or convex. However we cannot connect a concave and a convex parts in one solution. Considering this solution with two free edges, it is not suitable to describe the shape of closed vesicles. However, we can find two edges on it to satisfy the boundary conditions for open vesicles.

This work is supported by the National Natural Science Foundation of
China Grants 11304383.

\end{document}